\documentclass{ecai}
\usepackage{times}
\usepackage{latexsym}
\usepackage[hyphens]{url}  
\usepackage{graphicx} 
\usepackage{graphicx}  
\usepackage{amsmath,amssymb,amsfonts}
\usepackage[norelsize]{algorithm2e}
\usepackage{subfig}
\usepackage{commath}
\usepackage[noend]{algpseudocode}
\DeclareMathOperator*{\argmin}{argmin}
\usepackage{mathtools}
\usepackage{comment}
\usepackage{balance}
\usepackage{amsthm}
\usepackage{xcolor}
\newtheorem{definition}{Definition}

\ecaisubmission   

\begin{document}

\title{Anonymizing Data for Privacy-Preserving Federated Learning}

\author{Olivia Choudhury\institute{IBM Research Cambridge, email: olivia.choudhury1@ibm.com} \and Aris Gkoulalas-Divanis\institute{IBM Watson Health, Cambridge, email: gkoulala@us.ibm.com}  \and Theodoros Salonidis\institute{IBM T. J. Watson Research Center, email: tsaloni@us.ibm.com} \and \\ Issa Sylla\institute{IBM Research Cambridge, email: issa.sylla@ibm.com} \and Yoonyoung Park\institute{IBM Research Cambridge, email: yoonyoung.park@ibm.com} \and Grace Hsu\institute{Massachusetts Institute of Technology, email: ghsu@mit.edu. Research done during internship at IBM Research Cambridge} \and Amar Das\institute{IBM Research Cambridge, email: amardas@us.ibm.com}}

\maketitle
\bibliographystyle{ecai}

\begin{abstract}

Federated learning enables training a global machine learning model from data distributed across multiple sites, without having to move the data. This is particularly relevant in healthcare applications, where data is rife with personal, highly-sensitive information, and data analysis methods must provably comply with regulatory guidelines. Although federated learning prevents sharing raw data, it is still possible to launch privacy attacks on the model parameters that are exposed during the training process, or on the generated machine learning model. In this paper, we propose the first syntactic approach for offering privacy in the context of federated learning. Unlike the state-of-the-art differential privacy-based frameworks, our approach aims to maximize utility or model performance, while supporting a defensible level of privacy, as demanded by GDPR and HIPAA. We perform a comprehensive empirical evaluation on two important problems in the healthcare domain, using real-world electronic health data of 1 million patients. The results demonstrate the effectiveness of our approach in achieving high model performance, while offering the desired level of privacy. Through comparative studies, we also show that, for varying datasets, experimental setups, and privacy budgets, our approach offers higher model performance than differential privacy-based techniques in federated learning.

\end{abstract}

\section{Introduction}

Machine learning models often face significant challenges when applied to large-scale, real-world data. These may include decentralized data storage, cost of creating and maintaining a central data repository, high latency in migrating data to the repository, single point of failure, and data privacy. Federated learning (FL)~\cite{mcmahan2016communication} offers a new paradigm for iteratively training machine learning models using distributed data. At each iteration, the sites train a global model on their local data, typically using gradient descent method. The parameter updates of the local models are subsequently sent to an aggregation server and incorporated into the global model. The updated global model is again shared with the sites for the next iteration of training. The merit of this approach has been demonstrated in several real-world applications, including image classification~\cite{wang2019adaptive}, language modeling~\cite{mcmahan2016communication}, and healthcare~\cite{brisimi2018federated,choudhuryAMIA}. 

FL is particularly applicable in the healthcare domain, where data is rife with personal, highly-sensitive information, and data analysis methods must conform to regulatory requirements. Although FL is considered to be a step closer to protecting data privacy, it can still be vulnerable to various inference or poisoning attacks~\cite{bagdasaryan2018backdoor,wang2019beyond}. For instance, by initiating membership inference attack, adversaries can infer if an individual's data was used for training the model~\cite{shokri2017membership}. In reconstruction attack, adversaries aim to reconstruct the training dataset from model parameters~\cite{abadi2016deep,fredrikson2015model}.

A majority of recent work have adopted $\epsilon$-differential privacy (DP)~\cite{dwork2006our} for protecting FL models against such attacks. Although DP is considered state-of-the-art for offering strong privacy guarantees, in practice, it often yields low data utility due to the addition of excessive noise. As noted in~\cite{ choudhury2019differential,geyer2017differentially}, integrating DP with FL causes a significant reduction in data utility, particularly for a setup comprising less than 1000 sites. More importantly, the interpretation of the privacy parameter $\epsilon$, which restricts the impact an individual record has on the output of analysis, does not provide an intuition regarding what information is leaked about an individual~\cite{clifton}. Also, a given value of $\epsilon$ does not offer the same level of privacy across different datasets. As such, it is challenging to use DP for proving compliance on de-identification with privacy legal frameworks, such as EU General Data Protection Regulation (GDPR)\footnote[1]{GDPR law: \url{https://eur-lex.europa.eu/legal-content/EN/TXT/?uri=OJ:L:2016:119:TOC}} and the US Health Insurance Portability and Accountability Act (HIPAA)\footnote[2]{HIPAA law: \url{https://www.hhs.gov/hipaa/for-professionals/privacy/special-topics/de-identification/index.html}}, a crucial requirement in healthcare applications.

In this paper, we propose a syntactic approach for offering privacy in the context of FL. Unlike DP-based frameworks, our approach aims to maximize data utility and model performance, while enabling a provable and defensible level of privacy that adheres to the demands of privacy legal frameworks. Syntactic anonymity approaches support universal privacy guarantees that are interpretable. This has allowed policy makers to accept them as the standard for data protection. Moreover, syntactic approaches come with established processes for deciding on an acceptable level of privacy, given the data characteristics, intended use, security and contractual controls that are in place (c.f.~\cite{khaled} for de-identification under US HIPAA and Canada's PIPEDA, as well as guidelines from the Spanish Data Protection Authority on the use of syntactic approaches for GDPR anonymization~\cite{spainDPA}).

Although syntactic approaches have been studied in centralized settings, their potential has not yet been explored in a FL setup. Application of a syntactic approach in FL poses several challenges, which stem from the need to coordinate anonymization of data across sites, both during and after FL training. The first core component of our approach is to augment the FL training procedure with a syntactic anonymization step at the local sites. Specifically, we employ an anonymize-and-mine approach~\cite{charubook}, where we apply a syntactic anonymization method on the original private data and use the resulting anonymized data for subsequent mining. Our anonymization approach operates on data records that consist of a relational part and a transactional part, offering protection against adversaries who may have knowledge about individuals that spans these two data types. The second core component is a global anonymization mapping process that aids the resulting FL global model in the prediction process.  
We perform a comprehensive empirical evaluation of our approach on two important machine learning applications in the healthcare domain, using real-world electronic health data of 1 million patients. The results demonstrate the effectiveness of our approach in achieving high model performance, while offering sufficient privacy. We also provide a comparative analysis with DP, in terms of data utility, for various values of privacy parameters $k$ and $\epsilon$, commonly used in practice. Compared to DP, our approach achieves significantly better utility preservation and model performance and is more interpretable at the level of privacy it offers.

\centerline{}
\noindent The key contributions of this work include:
\begin{enumerate}
    \item Presenting the first syntactic approach to protect privacy in the context of FL. Applying a syntactic approach to FL is challenging because data is distributed among sites, and requires several novel steps beyond the existing centralized approaches. 
    \item Evaluating the proposed approach on two important problems in the healthcare domain, using two real-world large-scale health datasets.
    \item Comparing and contrasting $\epsilon$-differential privacy and our syntactic approach in the context of FL, with respect to the level of utility and interpretability, for typically-used privacy thresholds.
\end{enumerate}


\section{Related Work}\label{Related Work}
In this section, we review different privacy attacks on FL and summarize the state-of-the-art approaches to combat them. 


\vspace{2mm}
\noindent{\bf Privacy attacks on federated learning:}  
Although FL evades the need for sharing raw data, recent studies have identified potential privacy attacks that can still compromise the integrity of the model and data. FL is susceptible to privacy attacks at non-trusted sites and non-trusted aggregation servers, which can be broadly categorized into inference attacks and poisoning attacks. As described in~\cite{Nasr19AComprehensive} and references therein, inference attacks include membership attacks (infer whether a participant's record was used in the training dataset) and reconstruction attacks (infer the training dataset from model parameters). Inference attacks can be further categorized into black box (by accessing only model predictions) and white box (by accessing the model's parameters in addition to model predictions) attacks. In~\cite{Nasr19AComprehensive}, the authors proposed a white box inference attack, where users exploit the privacy vulnerabilities of stochastic gradient descent (SGD) algorithm in FL. Further, in~\cite{wang2019beyond}, the authors considered a FL scenario that experiences user-level privacy leakage due to attack from a malicious aggregation server. They proposed a reconstruction attack mechanism based on generative adversarial network (GAN). A more recent form of attack is poisoning attack, where users can manipulate parameters of their FL model updates in order to \textit{poison} the overall FL process in their desirable ways~\cite{bagdasaryan2018backdoor}. 

\vspace{2mm}
\noindent{\bf Privacy-preserving federated learning approaches:} 
Existing literature on privacy-preserving FL has primarily focused on DP and secure multiparty computation (SMC). SMC for FL has been proposed to compute sums of model parameter updates from individual user's devices in a secure manner~\cite{bonawitz2017practical}. Such an approach is applicable to specific operations, such as sum-based aggregation, and can protect from non-trusted aggregation server, but are computationally expensive in practice. Our proposed approach addresses the scenario of non-trusted servers, in addition to non-trusted sites, using lightweight computations, since each site only shares model parameters trained on anonymized data. 

Several recent approaches have proposed DP for FL using different implementation techniques~\cite{Agarwal18cpSGD,McMahan18AGeneral,Thakkar19Differentially}. Most of these approaches have focused on mobile application scenarios, such as image recognition or next-word predictor for mobile keyboards. They assume  a very large number of users, typically mobile phones, and focus on deep neural network models, which assume the existence of large-scale training data at sites. In addition, the proposed techniques improve model performance by exploiting the massive number of FL sites. The work in~\cite{geyer2017differentially} proposed a DP-based approach for health applications, but did not consider the scenario of non-trusted servers. In addition, the reported evaluations were not performed on real-world data. Prior studies have shown that utility or model performance can only be preserved for a setup comprising large number of sites (in the order of 1000 sites), and takes a severe hit when there exist fewer sites (in the order of 100 sites). The work in~\cite{Truex18AHybrid} implemented a mechanism combining SMC and DP. However, none of the above-mentioned work on DP for FL can provably achieve compliance with GDPR and HIPAA regulations around data de-identification and anonymization. 

In this paper, we focus on health applications and inference attacks that can be launched by sites as well as the aggregation server. These applications also do not assume a massive number of sites (sites are typically hospitals or healthcare institutes) and each site may not host large-scale data for deep learning models to be applicable. Such a scenario further necessitates the need to derive insight from other sites, in the form of FL, to construct more accurate models. To the best of our knowledge, this is the first work to propose a syntactic privacy-preserving mechanism for FL, which offers increased model performance and compliance with legislative frameworks, such as GDPR and HIPAA.
\begin{figure*}[!ht]
	\centering
	\includegraphics[width=.85\textwidth]{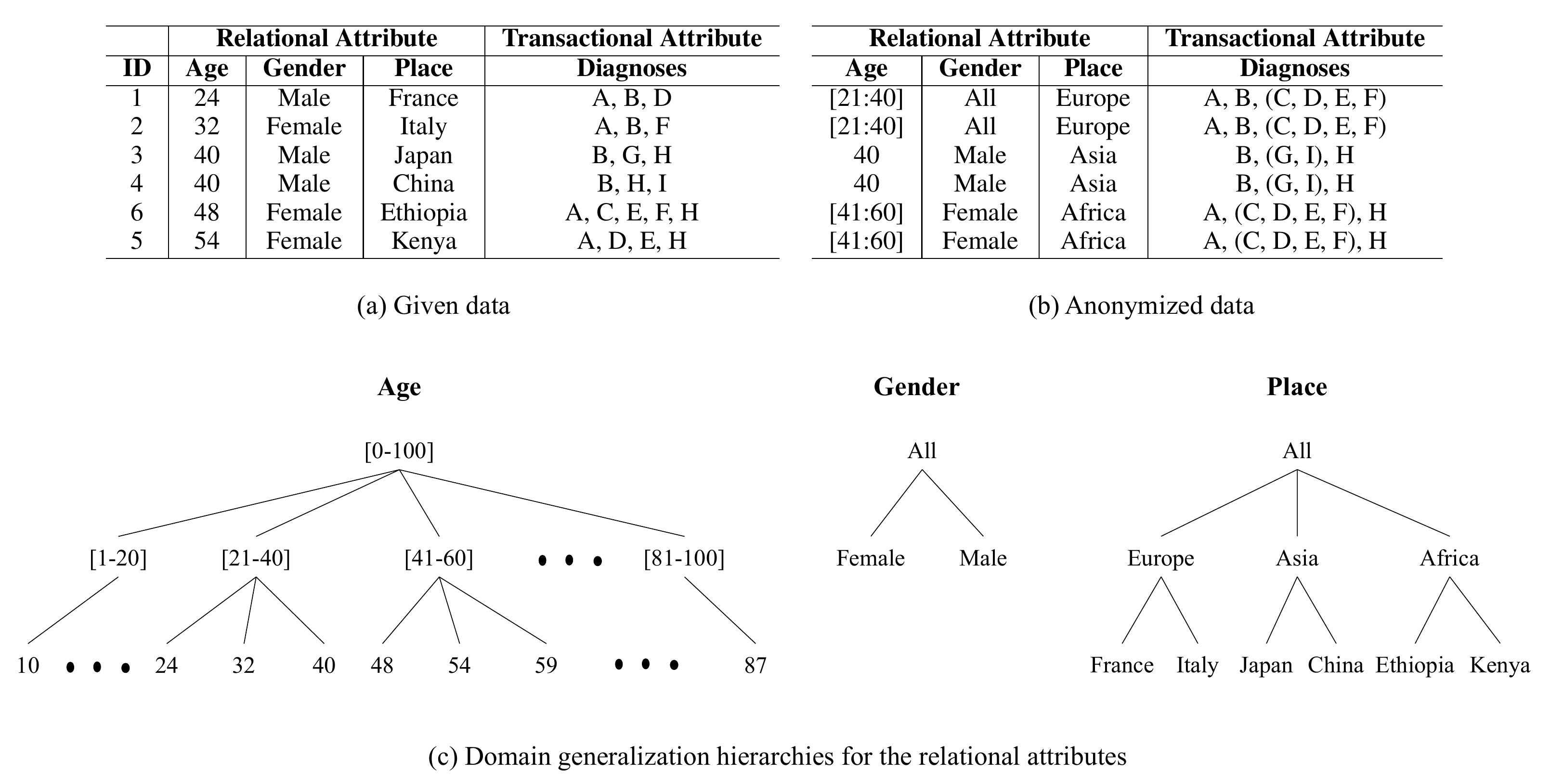}
	\caption{(a) $RT$-dataset containing relational (age, gender, place) attributes and transactional (diagnoses codes) attributes. (b) $(k, k^m)$-anonymized version of the data, where $k=2$ and $m=3$. (c) The hierarchy used to generate the anonymous data.} 
	\label{fig:MappingHierarchy}
\end{figure*} 



\section{A Syntactic Approach for Privacy in FL}\label{Methods}

In this section, we first provide the necessary background on syntactic approaches.  We then present our syntactic approach for FL. Specifically, we describe our proposed method for offering privacy in FL using a $(k, k^m)$-anonymity model. We identify the key challenges for adopting this approach in a FL scenario and propose solutions to mitigate them. We describe the underlying method through a series of steps, detailed in the sections that follow. 

\subsection{Background: Syntactic privacy models}\label{Background}
The notion of syntactic privacy was first introduced in the work of $k$-anonymity~\cite{sweeney2002k}. This data anonymization principle requires each record in a dataset to become indistinguishable from at least $k-1$ other records, with respect to the values of a set of potentially linkable attributes or \emph{quasi-identifiers} (QIDs). QIDs are attributes of a dataset that, in combination, can be used to re-identify individuals through triangulation attacks with other datasets. An example is the combination of date-of-birth, gender, and 5-digit zip code, which has been found to be unique for a large percent of US citizens.

\begin{definition}
Let $D(A_1,\ldots,A_u)$ be a relational dataset consisting of $u$ attributes, and $QID$ be a quasi-identifier associated with it. By construction, $QID$ involves a subset of attributes $A_r \subseteq \{A_1,\ldots,A_u$\}. Dataset $D$ satisfies $k$-anonymity with respect to $QID$, if and only if there exist at least $k$ records in $D$ for each sequence of values for attributes $A_r$.
\end{definition}

The dataset shown in Figure~\ref{fig:MappingHierarchy}(b), for example, is 2-anonymous with respect to QID attributes \texttt{age}, \texttt{gender}, and \texttt{place}, since every combination of the values of these attributes appears in (at least) $k=2$ records of the dataset.

$k$-anonymity has been widely adopted to preserve privacy of sensitive personal information, particularly in healthcare, marketing, and location-based services~\cite{gkoulalas2014publishing,niu2014achieving}. The $k$-anonymity approach has been further extended to other privacy formalism, such as $l$-diversity~\cite{machanavajjhala2006diversity} and $t$-closeness~\cite{li2007t}.

$k$-anonymity and its variants were designed for datasets containing only relational (numerical and categorical) attributes. In the context of healthcare, however, datasets also contain transactional (set-valued) attributes. In transactional datasets, individuals are associated with a number of items (known as an \emph{itemset}). Such items, for example, may be diagnosis codes, in which case an itemset is a set of diagnoses associated with a patient. To anonymize transactional data, the privacy principle of $k^m$-anonymity was invented~\cite{terrovitis2011local}.

\begin{definition}
Let $\mathcal{I}$ be the entire set of items that can be associated with a data record of a dataset $D$. Let $D$ be a transactional dataset over $\mathcal{I}$, where each record is associated with an item set $I \subseteq \mathcal{I}$. Dataset $D$ is $k^m$-anonymous if no attacker that has background knowledge of up to $m$ items of a record can use these items to identify less than $k$ records from $D$.
\end{definition}

The dataset shown in Figure~\ref{fig:MappingHierarchy}(b), for example, is $2^3$-anonymous with respect to the transactional attribute, as an attacker that has knowledge of any $m=3$ diagnoses associated with an individual (e.g., \texttt{A}, \texttt{B}, \texttt{D}), cannot use this knowledge to identify less than $k=2$ records from the dataset (in this example records with IDs 1, 2).

In modern datasets, individuals are typically associated with multiple types of data. In healthcare, for example, electronic health records involve both relational (numerical and categorical) and transactional attributes. In this context, relational attributes may correspond to patient demographics and transactional attributes to patient diagnoses. Anonymizing datasets that consist of both relational and transactional attributes (known as $RT$-datasets) is challenging due to the conflicting goals of minimizing information loss in relational and transaction attributes~\cite{poulis2013anonymizing}. This has led to the development of $(k, k^m)$-anonymization algorithms, which enforce $k$-anonymity on the relational attributes and $k^m$-anonymity~\cite{terrovitis2011local} on the transactional attributes~\cite{poulis2013anonymizing}. Since data anonymization unavoidably incurs data distortion, which leads to information loss, syntactic approaches that apply the $(k, k^m)$-anonymity principle offer privacy with bounded information loss ($\delta$) in one attribute type and minimal information loss in the other. 

\begin{definition}
Let $D(A_1,\ldots,A_u; B)$ be an $RT$-dataset consisting of $u$ relational attributes $A_1,\ldots,A_u$ and a transactional attribute $B$, and $QID$ be a quasi-identifier associated with the relational attributes. Dataset $D$ satisfies $(k, k^m)$-anonymity if no attacker that has background knowledge of the values of the quasi-identifier $QID$ for an individual and up to $m$ items of the transactional attribute $B$ associated with the same individual, can use this knowledge to identify less than $k$ records from $D$.
\end{definition}

In Figure~\ref{fig:MappingHierarchy}, we show an example of an $RT$-dataset and its $(k, k^m)$-anonymized counterpart that is built based on the given domain generalization hierarchies. Please observe that knowledge of the values of QID attributes \texttt{age}, \texttt{gender}, and \texttt{place}, as well as of up to $m=3$ diagnoses associated with a patient, always leads to (at least) $k=2$ records of the dataset. These records can have varying values for their non-identifying attributes.

Although syntactic approaches have been extensively addressed for centralized settings, they have not been considered in the distributed FL setting. 



\begin{figure*}[!ht]
	\centering
	\includegraphics[width=0.65\textwidth]{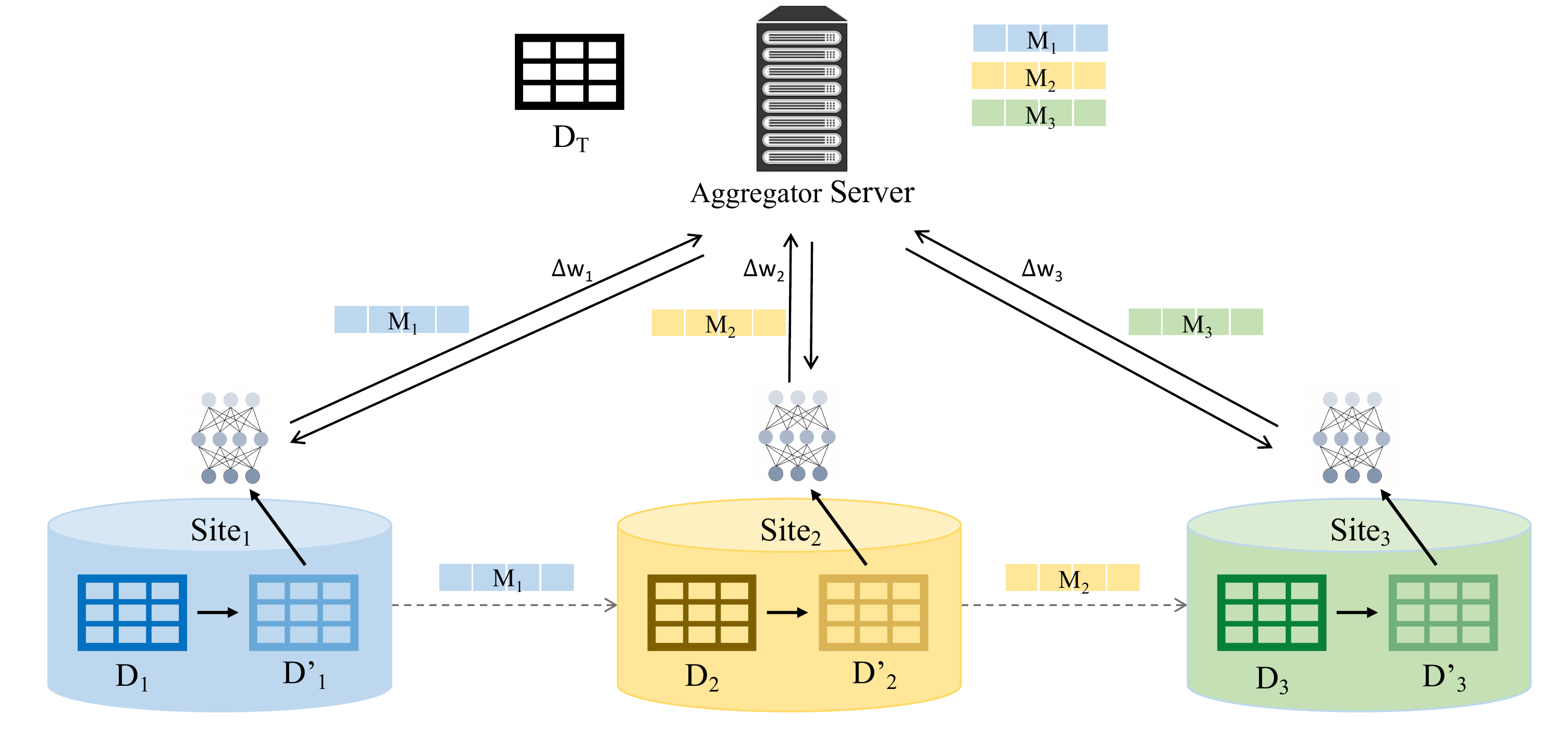}
	\caption{System design implementing our approach for privacy-preserving FL. The local data ($D_1, D_2, D_3$) at each site is anonymized using a syntactic approach. The syntactic mapping ($M_1, M_2, M_3$) generated at each site is shared with the aggregator server (or across sites) for future use. The anonymized local data ($D_1^{'}, D_2^{'}, D_3^{'}$) is used for training the federated model. When the aggregator server (or site) receives a new dataset ($D_T$), the samples are mapped to an appropriate equivalence class prior to using the federated model for predictive analysis.}
	\label{fig:SystemDesign}
\end{figure*}


\subsection{Selecting discriminative attributes}


At each site, we need to select the features or attributes to be used for training the model. Certain attributes, such as gender, date-of-birth, and zip code, of the local data qualify as QIDs which, however, may have low discriminative power for the classifier. Including such an attribute for training the model requires processing it as part of a QID and generalizing its values along with values of other attributes in the QID, to meet the $k$-anonymity requirement. This introduces noise to the data and often deteriorates the performance of the model. Hence, the first step of our proposed approach requires each site to determine the QID attributes that it should use for training its local model. Specifically, we rank the QID attributes based on feature importance to find the top discriminative ones and discard all others from the training of the local model. For our health datasets, we tested Recursive Feature Elimination (RFE), ExtraTreeClassifier and Random Forest techniques (RF) for computing feature importance at each site.

\subsection{Anonymizing local data}
\label{anonap}


The second step is to select an appropriate syntactic approach for anonymizing local data at each site. This selection needs to be done based on the types of attributes that exist in the dataset. Our health datasets contain both relational and transactional attributes, so we employ a $(k, k^m)$-anonymity-based approach~\cite{poulis2017anonymizing}.

We consider $N$ sites, each hosting its own local data $D_i$, where $i \in N$. Let $u_R()$ and $u_T()$ (these will be instantiated in section~\ref{expsetup} with equations \ref{n1}-\ref{u3}) be the functions measuring information loss for relational and transactional attributes, respectively. A lower value of these  metrics implies less information loss, hence better data utility. Furthermore, let $\delta$ be an upper bound of acceptable information loss in the relational data to accommodate for higher utility in the anonymization of the transaction data. Essentially, parameter $\delta$ aims to strike a balance between the conflicting goals of minimizing information loss in the relational data and minimizing information loss in the transactional data~\cite{poulis2017anonymizing}.

For a given dataset $\mathcal{D}$, we generate its $(k, k^m)$-anonymized version $\mathcal{D}'$, in a way that upper-bounds information loss in the relational part and minimizes information loss in transactional part. The anonymization is performed using the following three-step process.

\subsubsection{Original cluster formation}
In the cluster formation step, the algorithm produces $k$-anonymous clusters with respect to relational attributes only, in a way that aims to minimize information loss. Each record is represented as a multi-dimensional point, where each dimension corresponds to a QID attribute. A hard clustering (e.g., using an agglomerative method~\cite{kmeans}) is performed, where a cluster $S_j$ is formed for each set of at least $k$ points that are most similar with respect to their values for the QID attributes, using a data record similarity metric $u_R$. In the end of the clustering process, any ($< k$) points that have not been assigned to a cluster, are assigned to their closest cluster. Following that, for each formed cluster, the records corresponding to the points of the cluster are anonymized together by having their values for the QID attributes generalized to the same value. As an example, in Figure~\ref{fig:MappingHierarchy}(b), records with IDs 1, 2 are part of the same cluster and have to be anonymized together. For each QID attribute, the corresponding generalization hierarchy is used to locate the common ancestor of their values and use it to replace the original values (e.g., for attribute \texttt{age} the common ancestor of 24 and 32 in the Age hierarchy is [21-40], thus this value is used to generalize the records). In this way, a new dataset $\mathcal{D}_s$ is created that contains the generalized records from all clusters $S$.
    
\subsubsection{Iterative cluster merging} 
By construction, at the end of the cluster formation step, the identified clusters achieve minimal information loss with respect to the relational part. This comes at a cost of utility to the transactional part of the data. To reduce this effect and accommodate for lower information loss with respect to the transactional part of the data, we perform an iterative cluster merging process. This process aims to minimally reduce utility of the relational part ($u_R$) in an effort to significantly improve the utility of the transactional part ($u_T$), such that the $(k, k^m)$-anonymization solution retains acceptable utility in both parts of the data. To achieve that, we iteratively merge the set of clusters $S$ corresponding to $D_s$, to form larger clusters until we have reached the maximum allowable distortion of the relational part, as provided by parameter $\delta$. For cluster merging, we select a cluster $C$ as seed, from the list of clusters $S$, with minimum $u_R(C)$. We then create two orderings to sort the clusters in ascending order of $u_R$ and $u_T$. We select a cluster $C'$, such that it is closest to $C$ with respect to the two orderings and when merged with $C$, results in a dataset with $u_R$ satisfying $\delta$. Finally, we merge the clusters $C$ and $C'$ and update the corresponding records in $D_s$. The same process repeats as long as the produced clustering does not violate $\delta$. The final clustering that has not surpassed the maximum allowable distortion of the relational part is used in the next step to create the $(k, k^m)$-anonymized version of the dataset. For more details on iterative cluster merging techniques with similar objective, please refer to~\cite{poulis2013anonymizing}. In Figure~\ref{fig:MappingHierarchy}(b), iterative cluster merging results in three clusters: $c_1$ containing records with IDs 1, 2; $c_2$ containing records with IDs 3, 4; and $c_3$ containing records with IDs 5, 6.

\subsubsection{Enforcement of $(k, k^m)$-anonymization}
At this step, the final clusters have been formed and the records have been anonymized with respect to the relational part. To create a $(k, k^m)$-anonymized version of the dataset, we apply item generalization to the transactional attribute corresponding to the records of each cluster in $D_s$~\cite{terrovitis2011local}. Item generalization, illustrated in Figure~\ref{fig:MappingHierarchy}(b) with items placed inside parentheses, introduces uncertainty about which items of a generalized item are actually associated with the individual. For example, in Figure~\ref{fig:MappingHierarchy}, the generalized item (\texttt{G}, \texttt{I}) is interpreted as any (or both) of items \texttt{G} and \texttt{I} belonging to the record of the individual in the original dataset.

\subsection{Sharing of the syntactic mapping}
Given that all local models were trained using anonymous data records (generalized on their QID attributes to meet the requirements of the syntactic privacy model), the knowledge in the global model will be represented at the same aggregate level. Moreover, given that each site may have produced different generalizations of the QID attributes (e.g., due to the differences in the data distribution and number of records, or the value of $k$ used) to anonymize its data, the knowledge of the global model will span all such data generalizations. Let $M_i$ be the collection of all different combinations of values for the QID attributes (known as {\it equivalence classes}) that appear in the anonymized dataset of site $i$. In what follows, we use terms ``syntactic mapping'' and ``equivalence class'' interchangeably. Examples of equivalence classes (see Figure~\ref{fig:MappingHierarchy}) are: 
\begin{center}
    $M_1 \rightarrow$ Age : {\bf [21:40]}, ~~Gender : {\bf All}, ~~Place : {\bf Europe}, ~Diagnoses : {\bf A, B, (C, D, E, F)}\\
    $M_2 \rightarrow$ Age : {\bf [41:60]}, Gender : {\bf Female}, Place : {\bf Africa}, Diagnoses : {\bf A, (C, D, E, F), H}
\end{center}

Let mapping $M$ be the union of all $M_i$ (equivalence classes) produced at the local sites. The global model will be able to process new data records after these are represented under one of the equivalence classes in $M$. Therefore, the site that will use the global model will need to have knowledge of the collection $M$ of all equivalence classes for all sites. Once each local dataset $\mathcal{D}_i$ is anonymized to $\mathcal{D}'_{i}$, we share the syntactic mapping ($M_i$), computed at site $i$, with the aggregator server. Alternatively, this information can be shared across sites through a secure protocol (see dotted lines in Figure \ref{fig:SystemDesign}). We note that sharing the equivalence classes produced at a node does not violate privacy because for each equivalence class (by construction) there exist at least $k$ unique records (individuals) with the same values of the QIDs. No records are shared among equivalence classes.

\subsection{Training the FL model on anonymized data}
Following the norm of FL, we train and share a global model across all sites, using their syntactically anonymized data instead of their original data. We train the model based on the anonymized local datasets, after which the parameter updates are incorporated into the global model. This iterative process continues until the global model converges. For further details on training the FL model, see~\cite{choudhuryAMIA,mcmahan2016communication}. 

\subsection{Using the global FL model for predictions}\label{mapeqclass}

After training the FL model we can use it to perform predictions on new test data, which can be received at the server or at the local sites. 

The new data samples are in the form of the original data, while the FL model has been trained on anonymized data. As a result, we need to map each new sample to its most similar equivalence class from $M$, which is known to the global model. First, we select those mappings $M^*$ that are \textit{legitimate} for the data sample. A mapping is legitimate if, for each attribute of the equivalence class, the value of the sample for this attribute is the same or a subset of the corresponding value of the equivalence class. As an example, mapping $M_1$ is legitimate for data sample $t$, where
\{$t\rightarrow$ Age: {\bf 25}, Gender: {\bf Male}, Place: {\bf France}, Diagnoses: {\bf A}\}, while mapping $M_2$ is not, since -- for example -- \texttt{age} $25\notin [41:60]$. 

Among the legitimate mappings $M^*$ for $t$, we select the one that would require the least amount of generalization of the values in $t$ in order for $t$ to be placed under that equivalence class. We use $u_R$ and $u_T$ to calculate the information loss incurred for each relational and transactional attribute, respectively, to generalize it to the value of the corresponding attribute in the equivalence class, and take a weighted average to calculate the overall information loss. We assign $t$ to the mapping in $M^*$ that incurs the lowest information loss. As an example, assuming a mapping $M_3$:
\begin{center}
    $M_3 \rightarrow$ Age : {\bf [21:40]}, Gender : {\bf All}, Place : {\bf Europe}, Diagnoses : {\bf A}
\end{center}
we would select $M_3$ for transforming $t$ prior to providing it as input to the FL model for prediction, as it is more precise than $M_1$. 

In general, we select the mapping to use from $M^*$ by computing $\argmin\limits_j \{u_R(\mathcal{G}_R(\{t\cup M_j\})) + u_T(\mathcal{G}_T(\{t\cup M_j\}))\}$, where $\mathcal{G}_R$ denotes the generalization of data sample $t$ together with $M_j \in M^*$ across all relational attributes and $\mathcal{G}_T$ across all transactional attributes. As we will see in experimental section, the metrics $u_R$ and $u_T$ can also incorporate attribute weights that are set based on feature importance, penalizing more those mappings that incur significant information loss to highly discriminative attributes.  




\section{Experimental Evaluation}\label{Results}

In this section, we present a comprehensive evaluation of our method. We describe the real-world health data used in this study, followed by the experimental setup, and a comparative analysis.

\subsection{Use cases and data preparation}

Developing FL models and preserving their privacy are highly relevant in and applicable to the healthcare domain. To evaluate our proposed approach, we consider two important tasks for improving health outcome of patients: (a) prediction of adverse drug reaction, and (b) prediction of mortality rate. Adverse drug reaction (ADR) is a major cause of concern amongst medical practitioners, pharmaceutical industry, and healthcare system\footnote[3]{https://www.fda.gov/drugs/informationondrugs/ucm135151.htm}. As healthcare data is distributed across data silos, obtaining sufficiently large dataset to detect such rare events poses a challenge for centralized learning models. For the purpose of ADR prediction, we used Limited MarketScan Explorys Claims-EMR Data (LCED), which comprises administrative claims and electronic health records (EHRs) of over 3.7 million commercially insured patients. It consists of patient-level sensitive features, such as demographics, habits, diagnosis codes, outpatient prescription fills, laboratory results, and inpatient admission records. We selected patients who received a nonsteroidal anti-inflammatory drug (NSAID) to predict the development of peptic ulcer disease following the initiation of the drug. The selected cohort comprised 921,167 samples. We categorized demographic features (age, gender) as relational QIDs, and habits (alcohol, tobacco usage), diagnosis codes, and laboratory results as transactional QIDs. 

For the second use case, we considered the task of modeling in-hospital patient mortality. An accurate and timely prediction of this outcome, particularly for patients admitted to intensive care unit (ICU), can significantly improve quality of care. For this task, we used the Medical Information Mart for Intensive Care (MIMIC III) data~\cite{johnson2016mimic}. MIMIC III is a publicly available benchmark dataset, from where we derived multivariate time series from over $40,000$ ICU stays and labels to model mortality rate during ICU stays. As discussed in~\cite{harutyunyan2019multitask}, we selected $17$ physiological variables, including demographic details, each comprising $6$ different sample statistic features on $7$ different subsequences of a given time series, resulting in $714$ features per times series. The cohort consisted of $21,139$ ICU stays. We selected age, gender, height, and weight as relational QIDs.
   
\begin{figure*}[ht]
    \centering
    \subfloat[]{\includegraphics[width=0.25\textwidth]{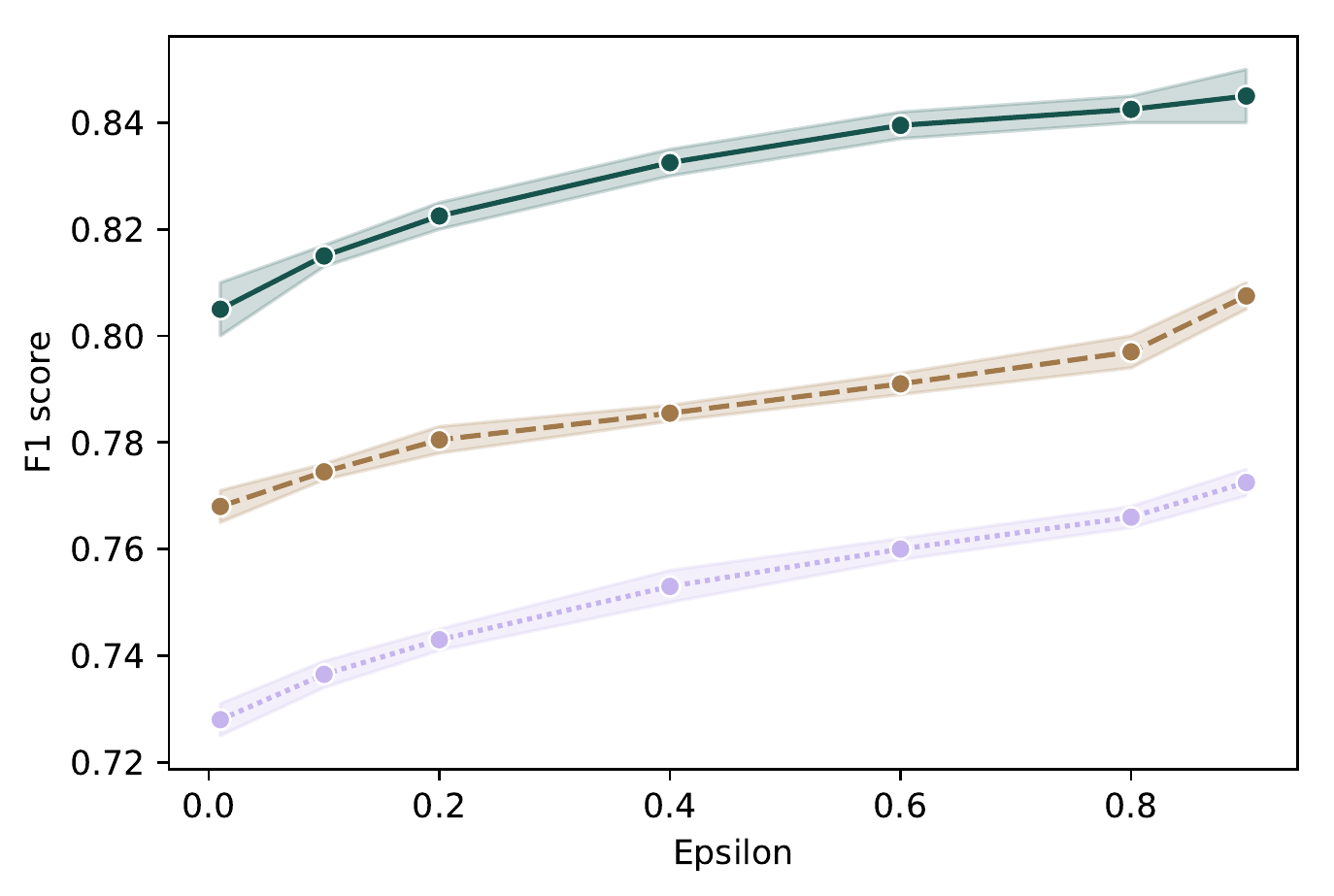}}
    \subfloat[]{\includegraphics[width=0.25\textwidth]{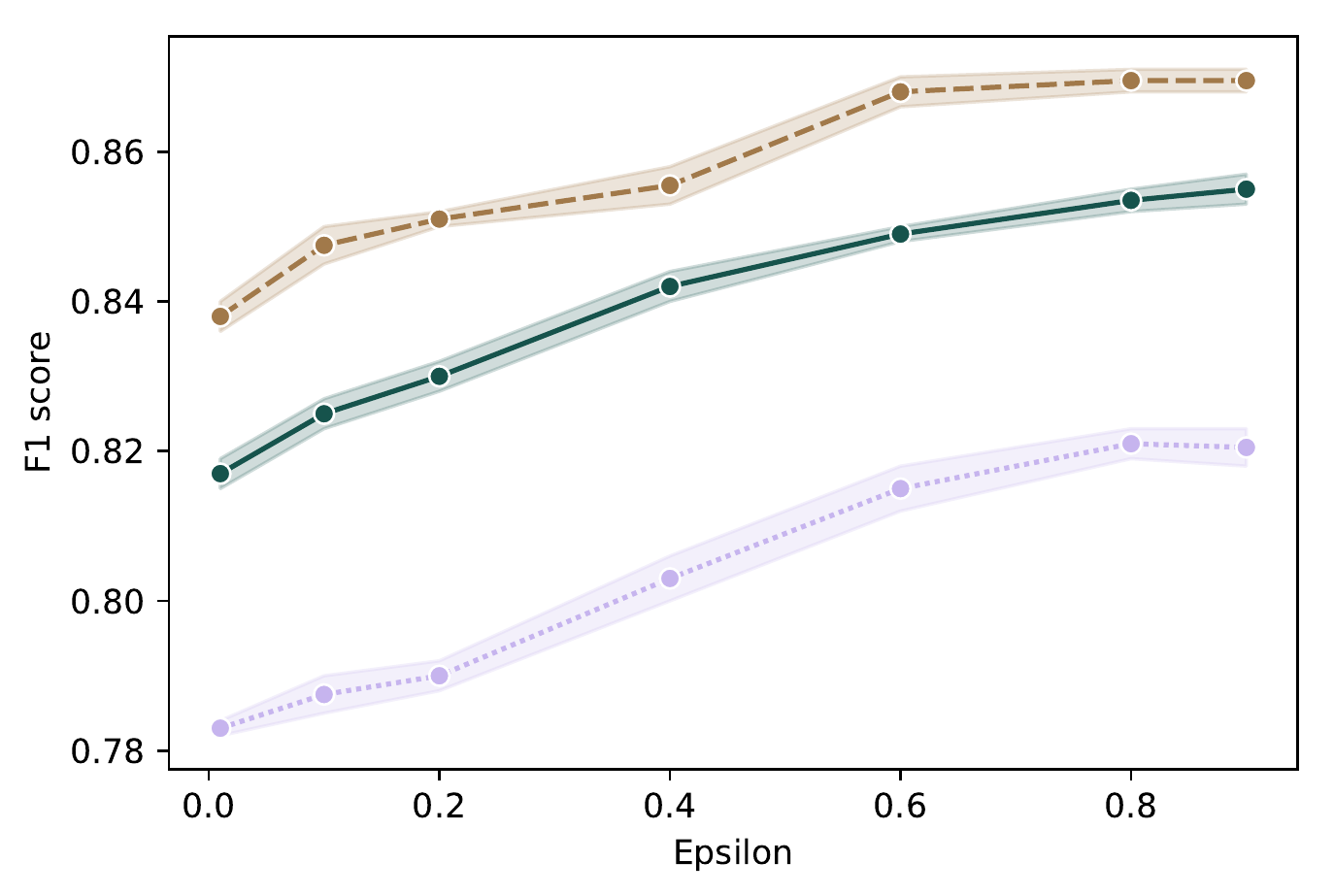}}
    \subfloat[]{\includegraphics[width=0.25\textwidth]{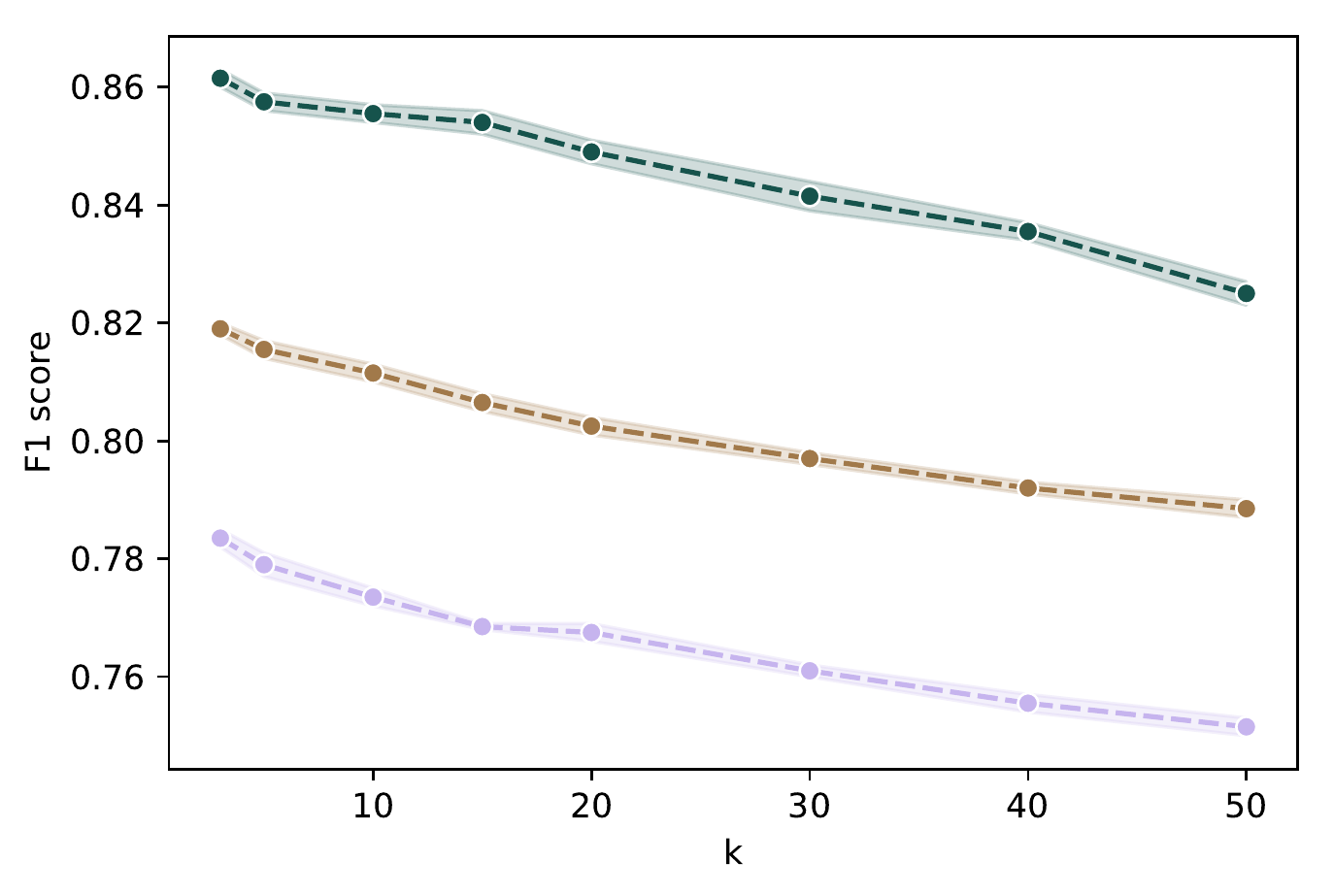}}
    \subfloat[]{\includegraphics[width=0.25\textwidth]{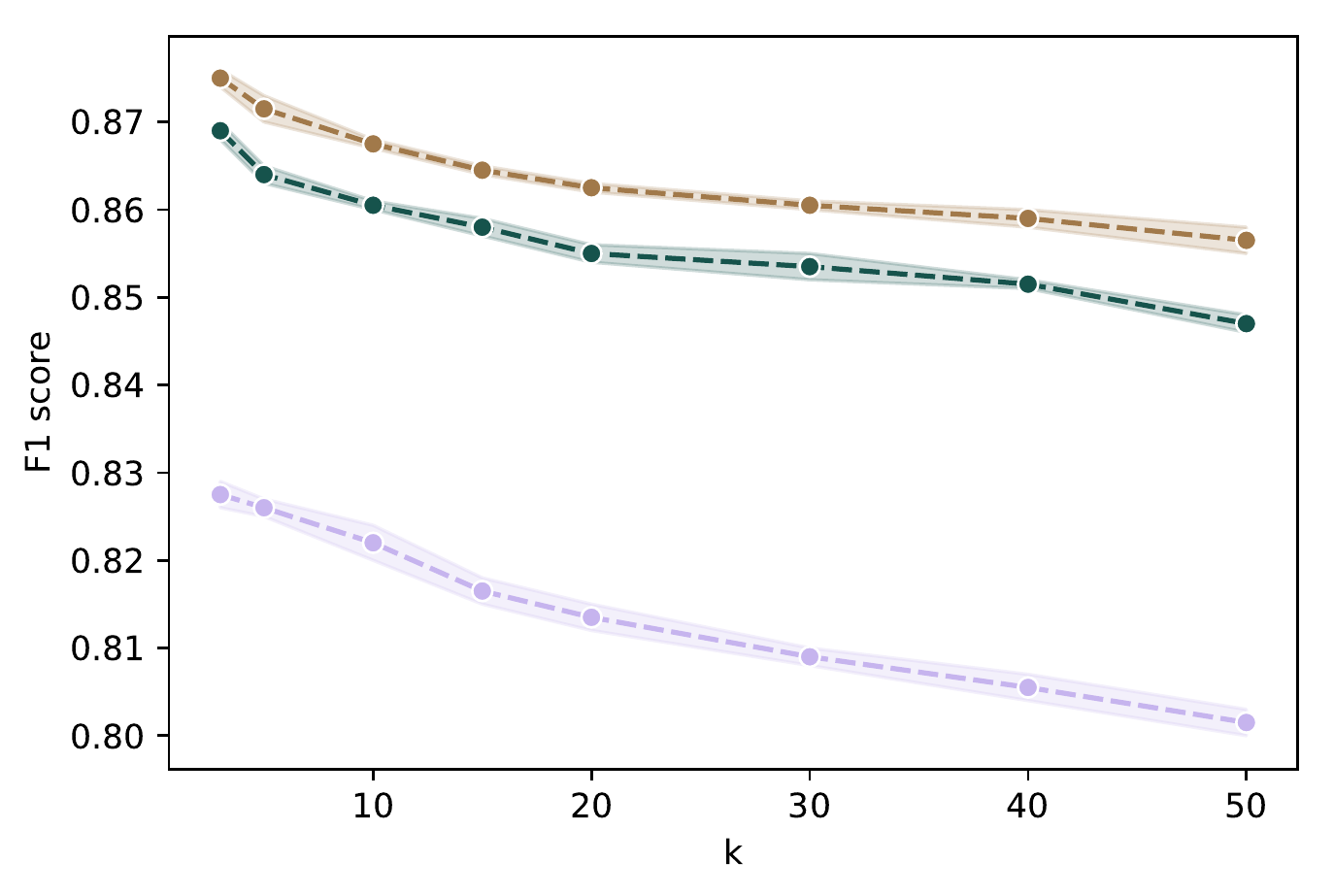}}    
    \caption{Effect of varying $\epsilon$ in $\epsilon$-differential privacy~\cite{chaudhuri2011differentially} for (a)LCED and (b)MIMIC data. The solid line, dashed line, and dotted line correspond to SVM, perceptron, and logistic regression, respectively. Effect of varying $k$ in our syntactic approach for (c) LCED and (d) MIMIC data. The dark (green) line, medium (brown) line, and light (mauve) line represent SVM, perceptron, and logistic regression, respectively.}
    \label{fig:Epsilon_f1_accuracy}
\end{figure*}

\begin{figure*}[h]
    \centering
    \subfloat[]{\includegraphics[width=0.39\textwidth]{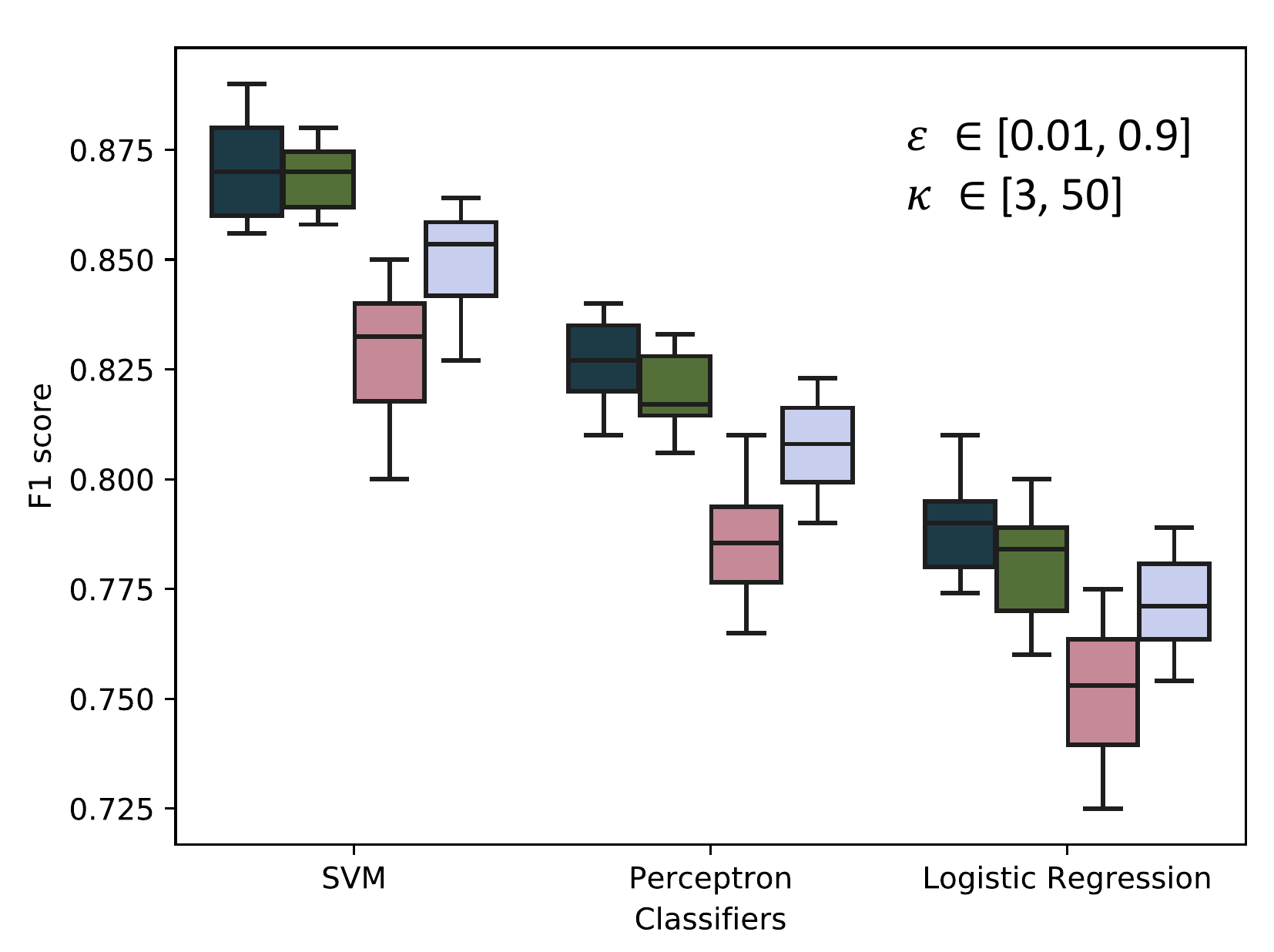}}
    \subfloat[]{\includegraphics[width=0.52\textwidth]{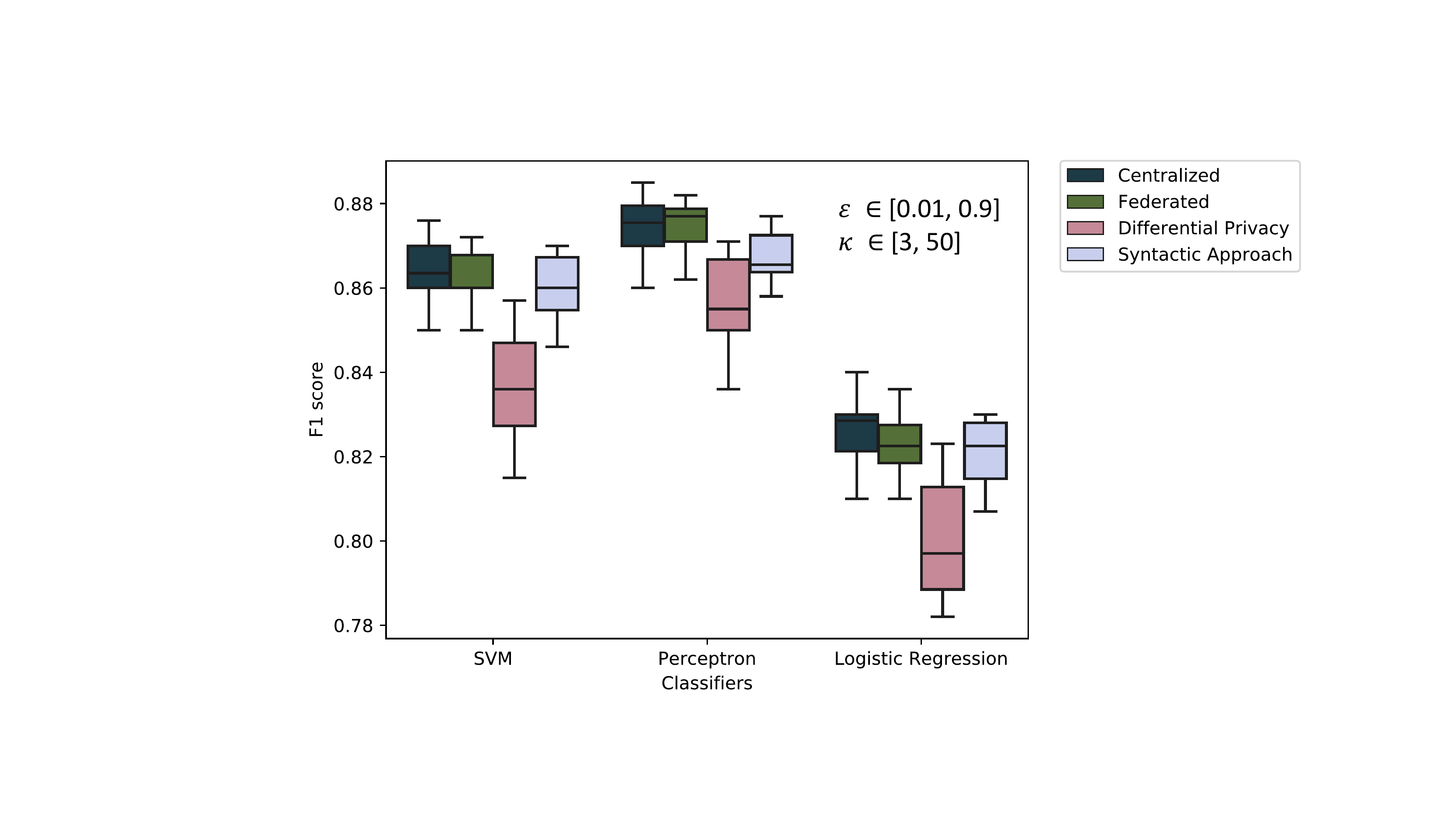}} 
    \caption{Comparison of F1 score with (a) LCED and (b) MIMIC data between centralized learning, FL, FL with $\epsilon$-differential privacy~\cite{chaudhuri2011differentially} ($\epsilon = [0.01,0.9]$), and FL with our proposed syntactic approach ($k = [3,50]$).}
    \label{fig:compare}
\end{figure*}   

\subsection{Experimental setup}\label{expsetup}
{\bf Machine learning algorithms.} To establish benchmark results, we first developed centralized learning models and FL models (with \emph{federated averaging}~\cite{choudhuryAMIA,mcmahan2016communication}) to predict ADR and ICU mortality. We used three classification algorithms, amenable to distributed solution using gradient descent, namely perceptron, support vector machine (SVM), and logistic regression. Logistic regression is widely adopted in the medical community for such tasks~\cite{harutyunyan2019multitask}, whereas SVM can handle highly imbalanced data~\cite{choudhuryAMIA}, which is typical in the ADR prediction use case. To evaluate the models, prior to and after employing privacy-preserving mechanisms, we measure their utility in terms of F1 score. 


\vspace{2mm}
\noindent {\bf Syntactic approach metrics.} For anonymization of data at local sites, we used the approach described in Section~\ref{anonap} with the metrics of normalized certainty penalty (NCP)~\cite{xu2006utility} for quantifying information loss due to generalization of relational attributes ($u_R$) and utility loss (UL)~\cite{loukides2011coat} for transactional attributes ($u_T$). NCP for a generalized value $\tilde{v}$, a record $r$, and an $RT$-dataset $\mathcal{D}$, is defined as: 
\begin{eqnarray}
NCP_R(\tilde{v}) = \begin{cases}
      0, & |\tilde{v}|=1 \\
      |\tilde{v}|/|R|, & \texttt{otherwise}
    \end{cases}\label{n1}\\
NCP(r) = \sum_{i\in[1, v]} w_i \cdot NCP_{R_i}(r[R_i])\label{n2}\\
NCP(\mathcal{D}) = \frac{\sum_{r\in\mathcal{D}}{NCP(r)}}{|\mathcal{D}|}\label{n3} 
\end{eqnarray}
, respectively, where $|R|$ denotes the domain size for a numerical attribute $R$ or the number of leaves in the hierarchy for a categorical attribute $R$, $|\tilde{v}|$ denotes the length of the range for a numerical $R$ or the number of leaves of the subtree rooted at $\tilde{v}$ in the hierarchy for a categorical $R$, and $w_i \in [0,1]$ is a weight to measure the importance of an attribute. The UL for a generalized item $\tilde{u}$, a record $r$, and an $RT$-dataset $\mathcal{D}$, is defined as: 
\begin{eqnarray}
UL(\tilde{u}) = (2^{|\tilde{u}|}-1) \cdot w(\tilde{u})\label{u1}\\
UL(r) = \frac{ \sum_{\forall \tilde{u}\in r} UL(\tilde{u})}{2^{\sigma(r)}-1}\label{u2}\\ UL(\mathcal{D}) = \frac{ \sum_{\forall r\in \mathcal{D}} UL(r)}{|\mathcal{D}|}\label{u3}
\end{eqnarray}
, respectively, where $|\tilde{u}|$ is the number of items mapped to $\tilde{u}$, $w(\tilde{u}) \in [0,1]$ is a weight to measure importance of $\tilde{u}$, and $\sigma(r)$ is the sum of sizes of all generalized items in $r$. The attribute weights $w$ in eq. (\ref{n2}) and (\ref{u1}) can be set using a feature importance computation method on each training dataset. Since we experimented and got similar results with Random Forests, ExtraTreeClassifier, and RFE (Recursive Feature Elimination with linear support vector classification as estimator and 50 features), we present the results of RFE. We set the privacy parameter $m=2$ and threshold $\delta=0.95$.

\vspace{2mm}
\noindent{\bf Comparative analysis.} For comparative analysis, we consider the state-of-the-art differential privacy mechanism~\cite{dwork2006calibrating,dwork2011firm,dwork2014algorithmic}. Differential privacy is a widely-used standard for privacy guarantee of algorithms operating on aggregated data. A randomized algorithm $\mathcal{A}(\mathcal{D})$ satisfies $\epsilon$-differential privacy if for all datasets $\mathcal{D}$ and $\mathcal{D'}$, that differ by a single record, and for all sets $\mathcal{S} \in \mathcal{R}$, where $\mathcal{R}$ is the range of $\mathcal{A}$,
\[
Pr[\mathcal{A}(\mathcal{D}) \in \mathcal{S}] \leq e^\epsilon Pr[\mathcal{A}(\mathcal{D'}) \in \mathcal{S}]
\] 

where $\epsilon$ is a privacy parameter. This implies that any single record in the dataset does not have a significant impact on the output of the algorithm. There are several methods for generating an approximation of $\mathcal{A}$ that satisfies differential privacy. We direct the readers to~\cite{geyer2017differentially,chaudhuri2011differentially} for details on implementing $\epsilon$-differential privacy in FL. 


We emphasize here that it is challenging to directly compare differential privacy and syntactic approaches due to the significant difference in their underlying notion. The privacy level offered by parameters $\epsilon$ and $k$ is not directly comparable. For this, in our experiments, we consider the range of F1 score for typical ranges of these parameters. This indicates the level of utility that the two approaches can support for an acceptable range of privacy. For the case of differential privacy, the range of values for $\epsilon$ were derived from state-of-the-art works in differential privacy~\cite{geyer2017differentially,chaudhuri2011differentially,nips2010oliver,nips2012hardt}. For our syntactic approach, the values of $k$ were selected following best practices described in~\cite{khaled} regarding values that have been used in practice across North America and Canada for data releases, compiled from several cases of data disclosures. These values span from $3$ to $20$, with the first being used for highly trusted data disclosures and the latter for highly non-trusted ones. To further evaluate the utility offered by our syntactic approach for even higher levels of privacy, we experimented with values of $k$ up to $50$.

\vspace{2mm}
\noindent {\bf Training setup.} All models were trained on 70\% of the data with 5-fold cross-validation. For FL, training data was randomly partitioned across 10 sites. Once trained, the model was tested using the 30\% test data. All experiments were run on an Intel(R) Xeon(R) E5-2683 v4 2.10 GHz CPU equipped with 16 cores and 64 GB of RAM. 

\subsection{Experimental results}


To establish benchmark results supported by $\epsilon$-differential privacy, we measure the privacy-utility trade-off for a given range of the privacy parameter. Figure~\ref{fig:Epsilon_f1_accuracy} (a) and (b) present the utility, measured by F1 score, for $\epsilon\in[0.01, 0.9]$ for the tasks of ADR and mortality prediction using LCED and MIMIC data, respectively. As $\epsilon$ increases, the level of privacy degrades, thereby improving the utility of the models. This is consistent across all three classification algorithms. 

We then evaluate our syntactic method in offering utility for a range of the privacy parameter $k$ that contains acceptable values for HIPAA and GDPR. Figures~\ref{fig:Epsilon_f1_accuracy} (c) and (d) show the variation of F1 score for different values of $k$. As the value of $k$ increases, more records in the dataset are generalized to form equivalence classes, which degrades the level of utility. This behavior is common in all three classification algorithms. 
Finally, we compare and contrast the performance of $\epsilon$-differential privacy and our proposed method in terms of utility, for the range of considered values of $\epsilon$ and $k$.  For a comprehensive study, we also compute the F1 score of centralized learning and FL. As shown in Figure~\ref{fig:compare}, our approach outperforms the state-of-the-art $\epsilon$-differential privacy method for all datasets and all classification algorithms. FL achieves comparable performance with respect to centralized learning with the additional benefit of not sharing raw data. As our approach ensures satisfying privacy while maximizing data utility, the predictive capability of the federated models coupled with the syntactic privacy-preserving approach is reasonable. However, the extent of performance degradation in FL when employing $\epsilon$-differential privacy is much severe.



\section{Conclusion}\label{Conclusion}
In this paper, we proposed the first syntactic anonymization approach for offering privacy in FL. Application of such an approach in FL is challenging due to the distributed source of training data, which requires several novel steps beyond a centralized anonymization approach: (a) deciding which quasi-identifiers to use at each site by considering the discriminative power of each feature along with data utility, to reduce the overhead of anonymization; (b) extracting and sharing syntactic mappings with the server; (c) transforming each test instance, using its most similar mapping, to the level of the data that have been used for training the global model. Our approach follows the anonymize-and-mine paradigm and operates on data records that consist of a relational and a transactional part. Through experimental evaluation on two real-world datasets and varying parameter settings, we demonstrated that our approach enables high model performance, while offering a defensible level of de-identification, as required by privacy legal frameworks.


\bibliography{ecai}
\end{document}